\newcommand{\kms}{{~\rm km\; s^{-1}}}

\newcommand{\msyr}{{~\rm M_{\odot}~yr^{-1}}}
\newcommand{\cm}{{~\rm cm}}
\newcommand{\s}{{~\rm s}}
\newcommand{\km}{{~\rm km}}
\newcommand{\g}{{~\rm g}}
\newcommand{\K}{{~\rm K}}
\newcommand{\erg}{{~\rm erg}}

\newcommand{\kpc}{{~\rm kpc}}

\newcommand{\AU}{{~\rm AU}}

\documentclass[12pt,preprint,a4paper]{aastex}
\usepackage{amsmath}                % American Mathematical Society package

\begin{document}

\title{GALACTIC VS. EXTRAGALACTIC ORIGIN OF THE PECULIAR TRANSIENT SCP~06F6}

\author{Noam Soker\altaffilmark{1}, Adam Frankowski\altaffilmark{1}, and Amit Kashi\altaffilmark{1}}

\altaffiltext{1}{Department of Physics,
Technion$-$Israel Institute of Technology, Haifa 32000, Israel;
soker@physics.technion.ac.il; adamf@physics.technion.ac.il; kashia@physics.technion.ac.il.}

\begin{abstract}
We study four
scenarios for the SCP~06F6 transient event that was
announced recently. Some of these were previously briefly discussed as plausible
models for SCP~06F6, in particular with the claimed detection of a $z=0.143$
cosmological redshift of a Swan spectrum of a carbon rich envelope.
We adopt this value of $z$ for extragalactic scenarios.
We cannot rule out any of these models, but can rank them from most to
least preferred.
Our favorite model is a tidal disruption of a CO white dwarf (WD) by an
intermediate-mass black hole (IMBH).
To account for the properties of the SCP~06F6 event, we have to assume
the presence of a strong disk wind that was not included in previous
numerical simulations.
If the IMBH is the central BH of a galaxy,
this explains the non
detection of a bright galaxy in the direction of SCP~06F6.
Our second favorite scenario is a type Ia-like SN that exploded inside
the dense wind of a carbon star.
The carbon star is the donor star of the exploded WD.
Our third favorite model is a Galactic source of an asteroid that collided with a WD.
Such a scenario was discussed in the past as the source of dusty disks around WDs, but
no predictions exist regarding the appearance of such an event.
Our least favorite model is of a core collapse SN.
The only way we can account for the properties of
SCP~06F6
with a core collapse SN is if we assume the
occurrence of a rare type of binary interaction.
\end{abstract}

%% {\bf Key words:}
%% \keywords{ \textbf }

% ====================
\section{INTRODUCTION}
% ====================

SCP~06F6 is an optical transient discovered by Barbary et al.
(2008). It raised to an optical peak flux of
$F_{\rm peak} \simeq 2.5 \times 10^{-14} \erg \s^{-1} \cm^{-2}$
in $\sim 100~$days, and then declined in another $\sim 100~$days.
Barbary et al. (2008) ruled out a microlensing event, and noted that
if SCP~06F6 is associated with the galaxy cluster CL 1432.5+3332.8 at a redshift of
$z=1.112$, then its peak luminosity is like that of the most bright SNe observed to date.
In this case, this object projected distance from the center of the cluster
would be $290 \kpc$, but
(1) there is no obvious host galaxy, making the
association questionable (a faint host is possible at a projected distance
of 12kpc for this redshift; Barbary et al. 2008).
(2) The spectrum of this object is not like any other SN.

Gaensicke et al. (2008a) presented a fit of the spectrum of SCP~06F6 with
a spectrum of a carbon-rich star at a redshift of $z=0.143$.
They also argue that the lightcurve looks somewhat like that of a supernova
type II.
Based on that they suggest SCP~06F6 represents a new class of
core collapse SNe.
However, this explanation shares the weak points (1) and (2) above, and in
addition:
(3) The X-ray luminosity reported by Gaensicke et al. (2008a) is two orders
of magnitude above that of a typical SN.
(4) The spectral fit is not perfect.

We discuss alternative possible scenarios for the SCP~06F6 transient.
In section 2 we describe the critical observations that set constraints on
any model to explain the transient.
In section 3 we discuss a Galactic WD-asteroid merger model.
Barbary et al. (2008) considered as well the possibility that the progenitor
was a cool WD in our galaxy.
In that case the WD progenitor should be quite cold, and its distance
should be $D>1.2 \kpc$, assuming the WD is not colder than $3000 \K$.
In section 4 we discuss three possible extragalactic models.
In addition to the core collapse SN model (Gaensicke et al. 2008a) we discuss
a type Ia-like SN scenario, where the exploding WD is enshrouded in a
carbon-rich nebula formed by the wind from the WD's giant companion.
We also discuss a possible disruption of a WD passing close to an
intermediate-mass black hole (IMBH; Rosswog et al. 2008b,c).
We compare the four models with observations and summarize in section 5.

% ======================================
\section{CRITICAL OBSERVATIONS}
% ======================================
\label{Sec_critical}

A model aimed at explaining SCP~06F6 should be able to simultaneously explain
its different observational features.
The crucial constraints are as follows
(see also Table 1).

(1) {\em The lightcurve.}
The lightcurve of SCP~06F6 is symmetric in time and akin to a bell curve.
The duration of the transient is $\sim 200$~days, and
only near the peak it resembles the lightcurve of a typical SN.
At later times SNe decline at
a much lower rate than the late evolution of SCP~06F6.

(2) {\em The spectrum.}
Barbary et al. (2008) checked that the SCP~06F6 spectrum does not resemble any known
SN spectra. As the closest, but not very convincing, matches they suggested
broad absorption line quasars or white dwarfs (WDs) with carbon lines (DQ WDs).
Indeed, the shapes and spacing of the absorption features in SCP~06F6 bear
resemblance to the C$_2$ Swan bands, present in DQ WDs and carbon stars.

Determination of temperature from the spectrum partially depends on
the object's galactic or extragalactic origin. Barbary et al. (2008)
estimated $T_{\rm eff} \ga 6500 \K$ for a galactic object from the slope
of the red continuum. Gaensicke et al. (2008a) hold that the presence of
Swan bands implies a temperature of 5000 -- 6000 K. However, Swan bands
are observed at higher temperatures, e.g. in DQ WDs at up to $\sim10,000 \K$
(Dufour et al. 2005). They generally weaken with increasing temperature,
but in a carbon-dominated atmosphere they would be extremely strong even
at 10,000 K (Dufour et al. 2008).

Barbary et al. (2008) find some evidence for spectral evolution
in the IR colors and the three optical spectra, but it is only mild.
The photospheric temperature remains apparently almost constant over
the 36 day period spanned by the three optical spectra.
The two main model constraints coming from the spectrum are a relatively
cool photosphere ($\sim 6000$K) and the presence of very wide bands, most
likely C$_2$ Swan bands.

(3) {\em X-ray emission.}
The X-ray flux was measured at one point in decline, about 80 days after maximum,
when the optical flux was at $\sim 20-30 \%$ of its peak value.
The X-ray flux was $\sim 4 $ times the peak optical flux, i.e.,
$\sim 10-20$ times the optical flux at the time.
It is quite possible that most of the emission from SCP~06F6 came in the X-ray band.

The presence of both X-rays and a photosphere at $T \la 6500 \K$
can be used to constrain the geometry of the event.
Most likely, the X-rays result from post-shocked gas that was ejected by the progenitor
at velocities of several$\times 1000 \km \s^{-1}$.
To preserve the photosphere at low temperatures, though,
the X-ray emitting gas has to be ejected over a small solid angle.
The most likely geometry is that of jets launched in the polar directions,
with the observer not along the polar axis; we rather face the disk sides.
Such a geometry indicates large specific angular
momentum, which in turn promotes models that involve binary systems,
e.g., merger, tidal disruption, or binary progenitor model.

% ======================================
\section{GALACTIC ORIGIN: A WD-ASTEROID COLLISION }
% ======================================

Although the galactic origin model is not our favorite one, we discuss it below for
three reasons.
(1) We cannot completely rule it out, and for that it serves us also in discussing
the other models.
(2) It might be applicable to other transient events in the future.
(3) It draws some parallel ingredients with the merger model of V838 Mon.
In addition, the process of WD-asteroid collision and the formation of
circum-WD material was discussed before in other contexts,
but no predictions exist regarding the appearance of such an event.

Some WDs are thought to possess dusty disks because of their
observed infrared excess (e.g. Zuckerman \& Becklin 1987; Graham et al.
1990; Becklin et al. 2005; Kilic et al. 2006; von Hippel et al. 2007).
The evidence was strengthened by the discovery that these WDs also
have metal-rich photospheres, presumably due to accretion of
metal-rich material (Koester et al. 1997; Zuckerman \& Reid 1998;
Jura et al. 2007; Zuckerman et al. 2007).
Even more recently, some hotter WDs have been found which exhibit
double-peaked Ca II emission lines while at the same time lacking
Balmer emission (Gaensicke et al 2006, 2007, 2008b).
The double-peaked Ca II emission profiles allow to conclude
that the metal-rich gas around these WDs has indeed disk-like geometry.
As discussed by Debes \& Sigurdsson (2002) and Jura (2003), such disks
can result from a tidal disruption of an asteroid or a comet colliding with
the WD. In this scenario, asteroids or comets are thought to be sent
on a collision course with the WD by some perturbation in the asteroid
belt or cometary cloud surrounding the WD.

% ======================================
\subsection{Model ingredients}
% ======================================
Consider the well studied transient event V838 Mon
(e.g., Wisniewski et al. 2008; Kami\'{n}ski 2008; Sparks et al. 2008;
and papers in  Corradi \& Munari 2007).
The most popular model for the eruption of V838 Mon is a merger event,
termed mergeburst, of two main sequence (or pre-MS)
stars (Soker \& Tylenda 2003, 2006; Tylenda 2005; Tylenda \& Soker 2006).
The basic ingredients of the merger model are as follows:
(1) Two objects collide, and the less dense object is completely destroyed (Shara 2002).
(2) The collision sometimes creates shockwaves that might lead to X-ray emission (Shara 2002).
(3) Some fraction of the mass of the less dense object is
accreted within a short time by the denser object.
This is the source of the large energy in the flash of the event.
(4) The other fraction of the mass of the destroyed objects, and possibly a
small amount of mass from the massive object, are ejected to large distances,
and form an extended envelope.
Some fraction of the extended envelope might escape the system.
(5) The release of gravitational energy continues as the inner parts of the
extended envelope contract and are accreted (Tylenda 2005).
(6) Because the release of the gravitational energy is due to a contracting envelope, the
energy release will continue as long as there is an envelope. For that, in the bright
phase we will not see the hot inner parts (assuming spherical symmetry; see below).
The merger will fade as a relatively cool object. This is in contrast to novae,
where the envelope disperses while the inner surface of the hot WD is still hot:
novae get bluer as they fade.

V838 Mon transient event is different in its spectrum, lightcurve, and
timescale from SCP~06F6.
However, we will examine what is required from a merging Galactic object to
explain an event like SCP~06F6.
Our proposed model is based on the same physical processes as in the model for
V838 Mon, replacing two MS stars with a large asteroid hitting a WD.

% ======================================
\subsection{Energy considerations}
% ======================================
Let the distance to SCP~06F6 be $D=10 D_{10} \kpc$.
Its peak luminosity is $L_{\rm p}=0.08 D_{10}^2 L_\odot$,
and the total radiated energy in the optical bands is (assuming the source is in our galaxy)
\begin{equation}
E_{\rm rad} \simeq 3 \times 10^{39} D_{10}^2  \erg
\label{eq:e1}
\end{equation}
For a black body temperature of $6500 \K$ the
photospheric radius of the object at peak luminosity is
\begin{equation}
R_{\rm ph} \simeq 0.2 D_{10} R_\odot.
\label{eq:rph1}
\end{equation}
Including the X-ray emission, the total radiated energy might be much larger, by
a factor of $\sim 5$ (Gaensicke et al. 2008a).

If the energy is released by accreting an amount of mass $m_a$, then the
energy released to radiation and to the inflated envelope is
$E=GM_{\rm WD}m_a/2R_{\rm WD}$.
Let a fraction $\eta$ of this energy be radiated in the optical
bands (by the photosphere). From their study of the stellar collision model
to the V838 Mon transient event, Soker \& Tylenda (2003) estimate that
$\eta\simeq 0.025-0.1$.
The rest of the gravitational energy will be released later, after the envelope
is dispersed.
The accreted mass based on the observations is then
$m_a \simeq  5 \times 10^{22} D_{10}^2 \eta^{-1} \g$,
where we took for the WD mass and radius $M_{\rm WD}=0.6 M_\odot$
and $R_{WD}=8,000 \km$, respectively.
If half the mass is accreted, and half resides in the envelope,
then the total mass of the object, an asteroid or a comet, is
\begin{equation}
m_2 \simeq  10^{24} D_{10}^2  \left( \frac{\eta}{0.1} \right)^{-1} \g.
\label{eq:m2}
\end{equation}
This is an asteroid with a diameter of $\sim 1000 \km$.

% ======================================
\subsection{Extended envelope structure}
% ======================================
The observations indicate a black body temperature of
$\sim 6500 \K$ (Barbary et al. 2008). We note that at this temperature the
opacity is $\kappa\sim 0.1 \cm^2 \g^{-1}$, and increases with temperature.
The dependence of opacity on temperature
in the $\sim 4,000-10,000 \K$ region is
$\kappa \simeq 10 (T/10^4 \K)^{10} \cm^2 \g^{-1}$
(Alexander \& Ferguson 1994; Marigo 2002; Ferguson et al. 2005).
The average density in the envelope is
$\bar \rho \simeq 3\times 10^{-8} D_{10}^{-1} (\eta/0.1)^{-1} \g \cm^{-3}$.
The optical depth is
$\tau \simeq \kappa R_{\rm ph} \bar \rho \simeq 50 (\eta/0.1)^{-1}$.
So an optically thick envelope is formed.

This actually sets the typical radius of the envelope for such kind of objects.
We demand an optical depth of $\tau \simeq 1$ at the photosphere,
where the density is much lower than the average density.
Therefore, the average optical depth of the envelope
will be much higher, and we take $\tau_e \sim 10$.
For an envelope mass of $5 \times 10^{23} (\eta/0.1)^{-1} \g$
and $\kappa=0.1 \cm^2 \g^{-1}$,
the condition $\tau_e \simeq \kappa R_{\rm ph} \bar \rho =10$ gives
$R_{\rm ph} =0.5 (\kappa/0.1 \cm^2 \g^{-1})^{1/2}
\eta^{-1/2} (\tau_e/10)^{-1/2} R_\odot$.

The effective temperature goes as $T_{\rm ph} \propto R_{\rm ph}^{-1/2} L_{ap}^{1/4}$,
where $L_{ap}$ is a possible alternative luminosity.
Using this expression in the dependence of opacity on temperature,
and then in the condition $\tau=\tau_e$, gives
\begin{equation}
R_{\rm ph} \simeq 0.3
\left(  \frac{\eta}{0.1} \right)^{-0.14}
\left( \frac{\tau_e}{10} \right)^{-0.14}
\left( \frac{L_{ap}}{0.08 L_\odot} \right)^{0.36} R_\odot.
\label{eq:rph2}
\end{equation}
The energy released is fixed by the asteroid that collided with the WD.
So the typical luminosity (about the peak luminosity) depends on the time duration
of burst $t_b$ as  $L_{ap} \propto t_b^{-1}$. Hence the radius depends on the duration
of the event as $R_{\rm ph} \propto t_b^{-0.36}$. Namely, if the event was longer
(shorter) by a factor of 10, the radius would be smaller (larger) by a factor
of $2.3$.
The dependence on the exact distribution of energy in the event, as expressed
by $\eta$, is also very weak.
Overall, we expect that for the WD and asteroid used here, the
photospheric radius would be $R_{\rm ph} \simeq 0.1-1 R_\odot$, with a very weak dependence
on the unknown variables.

The same arguments show that the steep increase of the opacity with temperature
means that there will be no strong temperature evolution with time.
Again, the reason is that as the envelope thins, the photosphere gets deeper to
hotter layers. But there the opacity increases, so the difference in the
physical depth into the envelope is not large.

The Kelvin-Helmholtz time of the envelope is
\begin{equation}
t_{\rm KH} \simeq \frac {GM_{WD}(0.5m_2)}{R_{\rm ph}} \frac{1}{L_{\rm p}}
\simeq 100 D_{10}^{-1} \left( \frac{\eta}{0.1} \right)^{-1} ~{\rm days}
\label{eq:tkh1}
\end{equation}

Overall, we can account for the optical lightcurve by emission from a semi-hydrostatic
envelope that is dispersed in a Kelvin-Helmholtz timescale. The envelope is not in
full hydrostatic equilibrium. If it deviates from a static configuration with a velocity
of $\sim 0.1$ times the photospheric sound speed, then the dispersion occurs over about one month.
Most of the envelope mass will be accreted onto the WD. But now, as we can see deeper into the
WD, and because of the angular momentum of the colliding asteroid, the system will
resemble a dwarf nova in an outburst.

At this stage we expect strong X-ray emission as well.
This is a possible explanation
for the strong X-ray emission from SCP~06F6 in the WD-asteroid merger model.
If indeed most of the emission comes in the X-ray band,
our explanation for that is the emission from near the surface of the WD.
As the thermal energy content of gas near the surface of the WD is larger,
the Kelvin-Helmholtz time will be somewhat longer for the X-ray emission
and the optical phase will be followed by an X-ray luminous phase.

The study of the X-ray emission requires 3D hydrodynamical simulation of the merger event.
However, we do note that accreting WDs can have a large fraction of their emission
in the X-rays.
For example, RU Peg has $\sim 1/3$ of its radiation coming in X-rays in quiescence,
when its accretion rate is $\sim 3 \times 10^{16} \g \s^{-1}$ (Silber et al. 1994).
The accretion rate during the decline of the WD-asteroid model is similar.
Some other WDs accreting in quiescence have their X-ray emission about equal
to that in the optical (van Teeseling et al. 1996).
The condition for that is that we see the WD surface. This can occur in
our merger model if after
the optical peak
the envelope becomes clumpy, or
because of the angular momentum of the asteroid, a very oblate atmosphere
is formed, which later flattens into a disk.

We note that equation (\ref{eq:tkh1}) ruled out a WD-WD merger.
A WD-WD merger was studied in the past in relation to the formation of
R Coronae Borealis (RCB) stars (e.g., Iben et al. 1996), and in
principle can be a source of a transient event.
However, for the luminosity and radius inferred
for SCP~06F6, the Kelvin-Helmholtz time is much too long.
Also, a non-exploding WD-WD merger cannot be an extragalactic source at
$z=0.143$, because the required luminosity would be much larger
than the Eddington luminosity of such a system.

% ======================================
\subsection{Spectrum}
% ======================================

\begin{figure}
\resizebox{0.7\hsize}{!}{\includegraphics{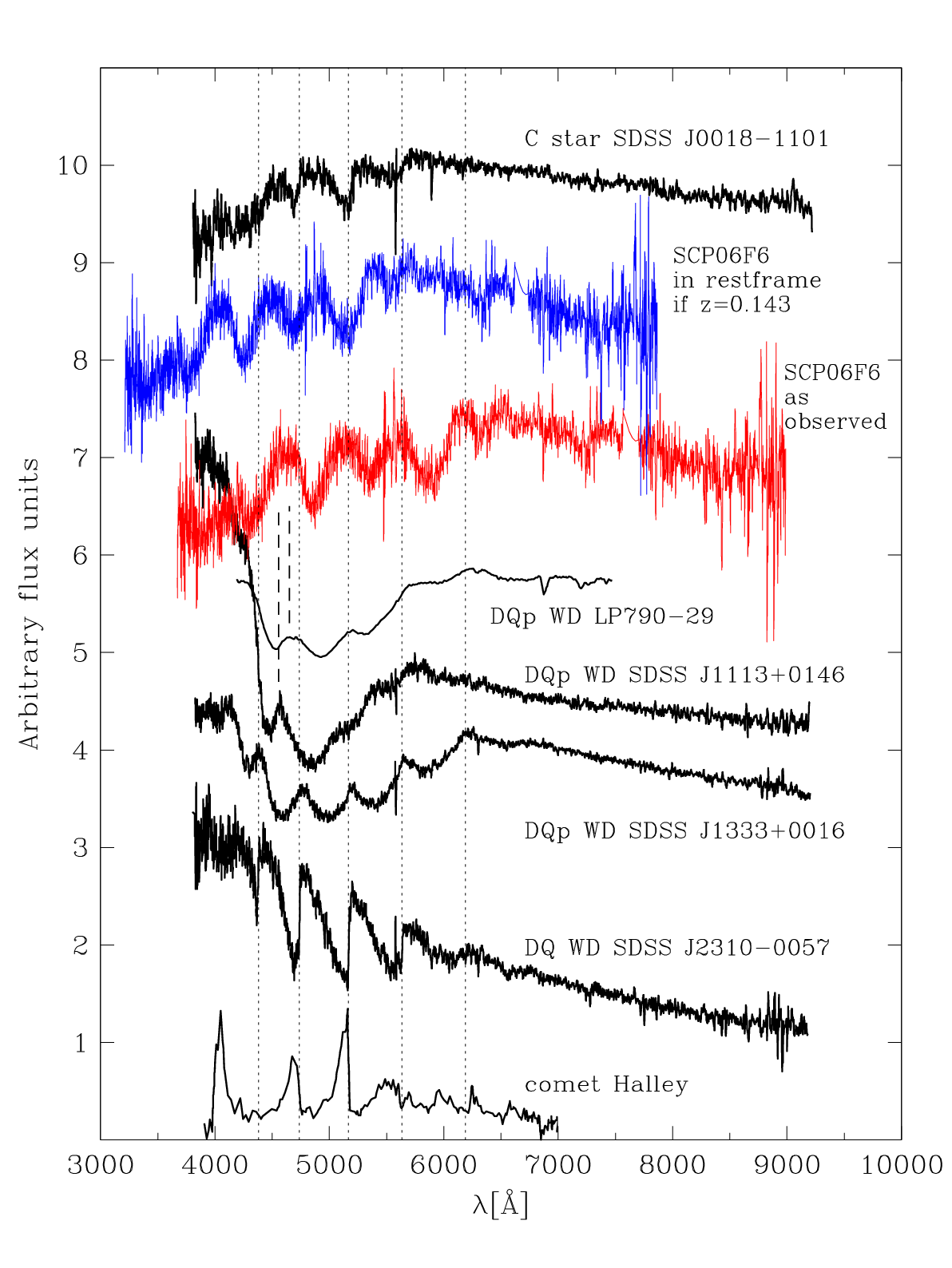}}
\caption{Comparison of the optical spectrum of SCP~06F6 (taken on 28.05.2006
with KECK, Barbary et al. 2008) to various objects exhibiting Swan bands.
The spectra, normalized to 1 at $6500$\AA\ and shifted vertically, are
labeled with the categories and names of the objects.
The top spectrum is of the same carbon star as found by Gaensicke et al.
(2008a) to be, when redshifted to z=0.143, the best match for SCP~06F6.
To assess their fit, this spectrum should be
compared to the second spectrum from the top, which is the spectrum
of SCP~06F6 in its
assumed restframe, if the transient is indeed located at z=0.143.
Dotted vertical lines mark band-heads of the C$_2$ Swan bands. Two short dashed
vertical lines connect the features of the two DQp WDs with an otherwise
unaccounted for maximum at 4500--$4700$\AA\ in the SCP~06F6 spectrum.
Note that the comet spectrum (bottom curve) features Swan bands
in emission. }
\label{Fig_spectra}
\end{figure}

In the WD-asteroid model the spectral features seen in SCP~06F6 are caused by the
C$_2$ Swan bands (or molecules containing also hydrogen) at $z=0$.
The observed optical spectrum is not easy to explain in this
case, as the bands observed in SCP~06F6 do not exhibit the typical
placement and shapes observed in Swan bands in most objects.
However, distortions and displacements of the C$_2$ Swan bands are
observed among the DQ WDs, especially in the peculiar DQp type
(Hall \& Maxwell 2008).
Their exact cause is still disputed, the main candidates being
magnetic fields or molecules related to, but different from C$_2$, such as
C$_2$H (see Schmidt et al. 1995, 1999; Hall \& Maxwell 2008).
Figure~\ref{Fig_spectra} compares the spectrum of SCP~06F6 taken by KECK
on 28.05.2006 (Barbary et al. 2008) with spectra of various objects
exhibiting Swan bands. With the exception of comet Halley
(de Freitas Pacheco et al. 1988) and LP 790-29 (Schmidt et al. 1995)
other spectra used for comparison are from the Sloan Digital Sky Survey
(SDSS, York et al. 2000; Adelman-McCarthy et al. 2008).

We note that the positions of the redder spectral
features of SCP~06F6 coincide reasonably well with the positions of normal
Swan bands (bandheads of the latter are marked by vertical dotted lines in
Fig.~\ref{Fig_spectra}), even though the shapes tend to be more symmetric
and rounded in the transient than in normal DQ WDs or carbon stars.
The general shape of the bands in SCP~06F0 is more similar to the
DQp stars than to the other objects featuring C$_2$.
To the blue, the broad maximum at 4500--$4700$\AA\ in SCP~06F6 is clearly
discrepant with the normal Swan bands and also with most DQp WDs.
It is only matched by a
corresponding maximum in the spectra of peculiar magnetic DQp WDs
LP 790-29 and  SDSS J1113+0146 (see the two short dashed vertical lines in
Fig.~\ref{Fig_spectra}).

A potential difficulty to the C$_2$ Swan band explanation of the SCP~06F6
spectrum is that normally the central bands are the strongest (usually the
$\Delta v = 0$ band with head at 5165\AA\ \!\!) and the bands to the sides
weaken gradually.
That is not the case for SCP~06F6, which has an
alternating weak--strong band intensity pattern.
In fact, this issue is not specific to the galactic model.
It is then fortunate that we can point to a magnetic
DQp WD with Swan bands and a similar
band intensity pattern -- namely G 99-37 (SDSS 0548+0010) which has
$B\sim10$MG (Angel 1977). In this case the alternating pattern results
from a presence of a strong CH band at 4300\AA\ in addition to the C$_2$
bands (e.g., Dufour et al. 2005).

There is no one ideal fit, but the best case seems to be LP 790-29, a
strongly magnetized WD with $B\sim 50$MG and
$T_{\rm eff}=7800$K (Liebert et al. 1978; Bues 1999).
The lack of a perfect match is not surprising as the temperature of
LP 790-29 is higher than that of the transient photosphere.
Also, the magnetic field in SCP~06F6 may be somewhat stronger, since its blue maximum
is more blueshifted than in LP 790-29, but less than in SDSS J1113+0146
(for which there is no $B$ determination).

Therefore, the variations in the C$_2$ bands observed among DQp WDs
could quite possibly account for the positioning of the SCP~06F6 features.
The requirement is that the magnetic field in the photosphere, at a radius
of $R_{\rm ph} \simeq 20 D_{10} R_{\rm WD}$, where $R_{\rm WD}$
is a typical radius
of a WD, be strong enough to modify the spectrum, i.e. $\sim100$MG.
If the magnetic field originates at the WD surface, then even assuming
magnetic field decreasing like
1/R$^2$, the field at the WD surface would have to be 40,000$D_{10}^2$MG.
This is a tall order, since the highest magnetic fields observed in WDs
are $\sim 1000$MG (Wickramasinghe \& Ferrario 2000; Vanlandingham et al.
2005; Kawka et al. 2007). Still, it can be fulfilled if the object is
relatively close, $D_{10}\la 0.16$, not much more distant than the lower
limit of 1.2kpc estimated by Barbary et al. (2008).
Magnetic field amplification in the accretion disk
may be another solution to account for the required high $B$.

Alternatively, some misalignments (most notably the peak around
4500--$4700$\AA\ )
can be understood if strong emission is present on top of strong
absorption bands.
Swan bands emissions are well known in comets (e.g., Krishna Swamy 1997),
they are also observed in R CrB stars in decline phase (Rao \& Lambert 1993).
Spectrum of comet Halley (from de Freitas Pacheco et al. 1988) is shown
as an example in Fig.~\ref{Fig_spectra}.
The cometary Swan bands emissions originate from resonant fluorescence
(Stawikowski \& Swings 1960) and their relative intensities vary between
comets (e.g., Fink \& Hicks 1996)
and within the same comet both spatially and temporally (O'Dell et al. 1988).
The hypothetical Swan band emissions in SCP~06F6 need not be the same as
in comets, but this example illustrates that some of the absorption features
could be strongly affected (displaced or even reversed) while others still
appear mostly unchanged.

Figure~\ref{Fig_spectra} presents also a comparison with the spectral fit
of Gaensicke et al. (2008a) who obtain $z=0.143$ for SCP~06F6.
The top curve displays their best matching carbon star spectrum,
SDSS J0018-1101.
Since all the comparison spectra in Fig.~\ref{Fig_spectra} are displayed
at z=0, assessing the fit of Gaensicke et al. (2008a) requires
also the transient spectrum to be transferred to its restframe.
The second curve from the top represents the restframe spectrum of
SCP~06F6, assuming the transient is indeed located at z=0.143.
It can be noted that their fit is appealing but also not ideal, with
each band having different deviations from their best case C star.

% ======================================
\section{EXTRAGALACTIC ORIGIN}
% ======================================
\label{Sect:extragalactic}

>From their spectral fit, Gaensicke et al.\ (2008a) determine a redshift
of $z=0.143$ for SCP~06F6 and, consequently, favor an extragalactic
explanation.
This redshift estimate being quite compelling, we adopt it
for our exploration of the extragalactic scenarios.
This sets immediate constraints for the energetics of the event and
the size of the observed photosphere.

The total peak observed flux derived from the $i_{775}$ band magnitude
and from the spectral energy distribution (IR/visible/UV) is
$2.5\times 10^{-14} \erg \s^{-1} \cm^{-2}$ (Barbary et al. 2008).
At a redshift of $z=0.143$
this translates to a peak luminosity of
$L_{\rm p} \simeq 1.3 \times 10^{42} \erg \s^{-1}$.
If the X-ray emission was in fact four times stronger, as it was later
at the time of the X-ray observation (Gaensicke et al. 2008a), then
$L_{\rm p,X}\simeq 5 \times 10^{42} \erg \s^{-1}$.
For a black body temperature $T_{\rm eff}$ and spherical geometry,
the radius of the photosphere at the peak luminosity is
\begin{equation}
R_{\rm ph-eg} \simeq
60\; \left(\frac{L_{\rm p}}{10^{42} \erg \s^{-1}}\right)^{0.5}\,
\left(\frac{T_{\rm eff}}{6500 \K}\right)^{-2} \AU .
\end{equation}
The lightcurve allows also a simple estimate of the total energy
radiated through the duration of the event, $t_{\rm event}$, as
$E_{\rm rad} \simeq L_{\rm p}\, t_{\rm event}/2
\simeq 10^{49} (L_{\rm p}/10^{42} \erg
\s^{-1}) \erg .$

% ======================================
\subsection{Supernova Models}
% ======================================

Both Barbary et al.\ (2008) and Gaensicke et al.\ (2008a)
consider SN origins for SCP~06F6.
While if at $z=1.112$ (as in Barbary et al.) the SN
would have to be among the most luminous observed just
to account for the optical flux, the $z=0.143$ value does
not require the conjectured SN to be overluminous.
Typical $L_{\rm p}$ of a supernova
is $10^{42}$--$10^{43} \erg \s^{-1}$, with core collapse SNe
being usually less bright but
exhibiting larger brightness diversity than thermonuclear ones
(e.g., Contardo et al. 2000; Richardson et al. 2002; Pastorello et al. 2005).
Typical radiated energy is $10^{49} \erg$.
Therefore the observed peak luminosity and energy budget of SCP~06F6
are consistent with the supernova hypothesis, be it thermonuclear
(like in SNe Ia) or core collapse.
SNe around maximum light exhibit spectral evolution on
a timescale of days to weeks, changing the appearance of spectral
lines and becoming gradually redder (e.g., Wells et al. 1994; Filippenko 1997).
Only slight spectral evolution was found by Barbary et al. (2008)
for SCP~06F6, but given the scarce spectral data for the transient,
there is no real inconsistency in this regard.

SNe are also known to be X-ray sources, but their X-ray
luminosities are 100 times lower than observed in this transient.
However, we note that the kinetic energy of SNe ejecta is
typically $\sim 10^{51}$~erg for both Ia and core collapse
SNe (e.g. Wheeler et al. 1995) which is
$\sim 100$ times larger than that observed as radiation.
Even higher kinetic energies are observed in the so called
hypernovae, which are defined by their kinetic energy exceeding
$10^{52} \erg$ (Paczy\'{n}ski 1998; Nomoto et al. 2006).
Through shocks formed by the ejecta colliding with the matter
surrounding a SN, a fraction of this kinetic energy can be converted
into X-ray emission: $\sim 1\%$ would be enough to explain the
observation of SCP~06F6.

The slow brightness rise, large photosphere size,
the Swan spectrum, and strong X-ray emission
set a general requirement for a supernova model:
a dense C-rich environment around the progenitor.
Let us estimate the density and mass in this envelope that
are required to form a photosphere at $R_{\rm ph-eg} \sim 60\AU$ .
The optical depth in the envelope is:
\begin{equation}
\tau = \int_{R_{\rm ph-eg}}^\infty \kappa \rho\, dr\; .
\end{equation}
For the photosphere $\tau = \frac{2}{3}$, and we will again use
$\kappa \sim 0.1 \cm^2 \g^{-1}$.
Assuming a 1/$R^2$ density profile for the wind, the density
of the envelope at the photospheric radius $R_{\rm ph-eg}$ becomes:
\begin{equation}
\rho\left(R_{\rm ph-eg}\right) \simeq 7.4\times 10^{-15}
\left(\frac{\kappa}{0.1 \cm^2 \g^{-1}}\right)^{-1}
\left(\frac{R_{\rm ph-eg}}{60 \AU}\right)^{-1} \g \cm^{-3},
\end{equation}
and the total mass of the envelope within radius $R_{\rm ph-eg}$ is:
\begin{equation}
M_{\rm en}\left(R_{\rm ph-eg}\right) \simeq 3\times 10^{-2}
\left(\frac{\kappa}{0.1 \cm^2 \g^{-1}}\right)^{-1}
\left(\frac{R_{\rm ph-eg}}{60 \AU}\right)^{2}
M_\odot .
\end{equation}

We can then estimate the wind mass loss rate required to get
a mass of $\simeq 0.03 M_\odot$ within the $60 \AU$ radius with a constant
wind velocity, $v_{\rm wind}$:
\begin{equation}
\dot M \simeq 10^{-3}
\left(\frac{\kappa}{0.1 \cm^2 \g^{-1}}\right)^{-1}
\left(\frac{R_{\rm ph-eg}}{60 \AU}\right)
\left(\frac{v_{\rm wind}}{10 \kms}\right) \msyr.
\label{Eq_mlr}
\end{equation}
If the dense wind is concentrated towards a preferred plane
and we see
the object roughly edge-on, as suggested by the existence of both
a cool photosphere and strong X-rays (see section~\ref{Sec_critical}),
the involved mass can be somewhat lower. An outflow limited to
$\pm 30^\circ$ from the plane would lower the required
envelope mass and mass loss rate by a factor of 2.

% ======================================
\subsubsection{The SN IIn connection}
% ======================================

It should be noted that there exists a subclass of SNe interpreted
as stellar explosions interacting with a dense circumstellar environment,
the so-called type IIn SNe (Schlegel 1990).
Their spectra and lightcurves are strongly influenced by the
circumstellar material (CSM) and some of them are strong radio and X-ray
sources.
It is noteworthy that of the two SNe that Gaensicke et al. (2008a)
bring as roughly matching the lightcurve of SCP~06F6, SN 1994Y is
type IIn (Ho et al. 2001) and SN 2006gy can be classified as a hybrid
type between Ia and IIn (Ofek et al. 2007).

SNe IIn are usually considered to be core collapse events,
but they are a very heterogeneous group (e.g., Filippenko 1997; Hoffman 2007)
and it was suggested that they can also originate from
thermonuclear SNe (Hamuy et al. 2003; Kotak et al. 2004; Aldering et al. 2006).
The first case suggested to be a thermonuclear SN IIn
was SN 2002ic, for which Hamuy et al. (2003) proposed
a scenario involving an AGB companion in a SN Ia-like system.
This SN became a prototype for the hybrid Ia/IIn type, also
nicknamed IIa (Deng et al. 2004; Aldering et al. 2006)

SNe IIn are characterized by narrow or moderately broad Balmer
emissions superimposed on the typical broad SN lines.
For example, a few days after maximum of SN 2002ic
mainly the narrow component of H$_{\alpha}$ line was visible
(Hamuy et al. 2003).
In this case, brightness in the
I band was almost constant during 60d after maximum, but
the lightcurve shapes among this group are very diverse.
Another type IIn, SN 1994W, exhibited a 120d plateau
after which the flux dropped suddenly, by 3.5 mag in the V band within
12 days and a similar amount in the R band
(Chugai et al 2004).

Typical estimated pre-SN mass loss rates of SN IIn are
in the range  $10^{-5}-10^{-3} \msyr$ (van Dyk et al. 1996),
while some of them are reported to reach even $\ga 10^{-1} \msyr$
for a short time before the explosion (Chugai \& Danziger 2003).
The winds of the suggested Ia/IIn keep to the extreme side:
$10^{-3}-10^{-2} \msyr$ (Hamuy et al. 2003; Trundle et al. 2008).
For the controversial case 2006gy (IIn or Ia/IIn) the immediate
pre-SN mass loss rate has been estimated at $\sim 10^{-1} \msyr$
(Ofek et al. 2007) or even $\sim 1 \msyr$ (Smith et al. 2008).
On average, SN IIn are almost as bright in the optical as type Ia
(Richardson et al. 2002)
and some recent examples are among the most energetic SNe ever observed --
most of these have been proposed to be Ia/IIn (e.g., Smith et al. 2008).
It has recently been argued that some core collapse SNe may explode
from Luminous Blue Variables (LBV) as progenitors (Kotak \& Vink 2006),
in particular for type IIn (Gal-Yam et al. 2007).
This led to the development of a modified SN-CSM interaction scenario
for the brightest type IIn SNe,
involving a very dense, {\em opaque} envelope close to the SN progenitor
(Smith \& McCray 2007; Smith et al. 2008), formed by the intense
LBV wind of the progenitor, most likely by a giant eruption of the
$\eta$ Car type (as opposed to a companion wind in the
SN Ia-like scenario of Hamuy et al. 2003).

SCP~06F6, if located at $z=0.143$, was optically bright, but
not overluminous, $M_I \sim -18$ (Gaensicke et al. 2008).
Its X-ray brightness, on the other hand, was very high.
A dense, optically thick C-rich circumstellar environment
is required to explain its characteristics and there might
be a hint of a narrow H$_\alpha$ line in the VLT and KECK
spectra (notice a small spike at $\sim 6560$\AA\ in the assumed
rest-frame spectrum of SCP~06F6 in Fig.~\ref{Fig_spectra}).
Therefore SCP~06F6 could be a case similar to type IIn SNe, but
with a C-rich envelope -- while for typical IIn or even for
Ia/IIn SNe the CSM is H-rich.

Whatever the origin of the envelope, it has been shown by
Smith \& McCray (2007) that when it is opaque, photon diffusion
of the thermal energy deposited in this envelope by shocked ejecta
is able to reproduce the rounded and broad maximum light of SN 2006gy.
This could be even more true for SCP~06F6, since the late rapid
decline of this photon diffusion model was actually too fast for
SN 2006gy --
additional energy sources had to be included for the late light,
which for SCP~06F6 would not be required.
In the Smith \& McCray (2007) scenario the faint X-ray emission of
SN 2006gy is attributed to a later interaction of the shock
with a less dense CSM, after the shock escapes the
opaque envelope. The extremely strong but transient X-rays from
SCP~06F6 differ from the SN 2006gy case and probably signify
a different structuring of the CSM, e.g., a bipolar configuration.

Other suggestions for the origin of SNe Ia/IIn include
pair instability SNe (Smith et al. 2007)
and Quark Novae (Leahy \& Ouyed 2008).
These could also be
tried for
SCP~06F6, but exploring the more exotic possibilies is
beyond the scope of the present paper.

% ======================================
\subsubsection{Core collapse SN}
% ======================================

A core collapse SN is the explanation favored for SCP~06F6 by
Gaensicke et al. (2008a). This is based on the spectral fit to
a carbon star spectrum that gives $z=0.143$, but the spectrum
itself is not explained, instead they assume it to be a new
class of SNe.
A basic problem with this scenario is that it
requires a young massive progenitor, of a main sequence
mass of $\ga 8 {\rm M_\odot}$, while there is no
bright star forming galaxy in the vicinity of SCP~06F6.
There is a possible host galaxy at $1.5\arcsec$ from the transient
position (Barbary et al. 2008) and at z=0.143, $1.5\arcsec$ is 3.7kpc.
But if located at that redshift, this galaxy would be very faint,
M${_Z}\sim -13.2$ (Gaensicke et al. 2008a).

Timescales of core collapse SN lightcurves are similar to that
observed in the transient, but typically
their declining parts are somewhat longer.
There are many core collapse SNe which declined $\sim 2$ mag in the
I band within 100d. It is harder to find examples of $\sim 3$ mag
decline within 100d. Except for some SNe~IIn, as explained above.

The dense C-rich circumstellar envelope required by a SN model could
be formed by a wind from the SN progenitor star itself or from its
hypothetical companion. A companion could also naturally explain bipolar
geometry of the matter. Winds from the canonical
core collapse progenitors, Wolf-Rayet (WR) stars,
can be dominated by carbon (the WC and WO types), but despite that
C$_2$ Swan bands were not observed in any SN.
WR winds are characterized by $v_{\rm wind} \ga 1000 \kms$ and
$\dot M_{\rm wind} \la 4 \times 10^{-5} \msyr$
(Crowther 2007).
Substituting these values in Eq.~(\ref{Eq_mlr}) indicates that the density
of a WR wind would be over 3 orders of magnitude lower than
required to form a photosphere at $60 \AU$.

LBVs undergo giant eruptions during which their
mass loss rate increases to $10^{-2} \msyr$ or even $\sim 1 \msyr$
(e.g., Smith \& Owocki 2006).
The high mass envelope could be provided by such an eruption shortly
preceding the SN event,
as suggested by Smith et al. (2007) for the proposed LBV
progenitors of SNe.
However, for SCP~06F6 a problem arises that LBVs not only
are not C-rich, but they are carbon depleted
(e.g., Garc\'{i}a-Segura et al. 1996).
Having {\em both} a WR and an erupting LBV in the system might provide
the needed total density of matter and right abundance of carbon,
but it would introduce an issue of wind interaction -- it seems that
the two winds would have to be mixed thoroughly to provide
a carbon-dominated spectrum from the photosphere, lasting at least
36 days. Also, a chance to encounter a system with appropriate timing
is small
(Smith 2008 estimates that it may happen in 0.05\% of all WR binaries).

Finally, after the core collapse explosion, a few solar masses of
SN ejecta moving with some $10^4 \kms$ would
sweep up the few times $10^{-2} M_\odot$ of the wind envelope
contained up to the $60 \AU$ radius in $\sim 10$ days, even if
the latter was concentrated around the equatorial/orbital plane.
Therefore a preexisting steady wind cannot
explain the relatively stable C-rich photosphere observed
in SCP~06F6 at $\sim 100$d from the start of the event.
Only a well timed giant LBV eruption could supply enough matter to
withstand the subsequent SN.

% ======================================
\subsubsection{Thermonuclear SN}
% ======================================

Another possibility is that SCP~06F6 was a SN Ia-like event
occuring within an envelope of carbon-rich matter, a thermonuclear
and C-rich version of SN Ia/IIn.
The SN progenitor WD needs to accrete mass from a companion to
exceed the Chandrasekhar mass.
In this case the C-rich envelope would be explained by a carbon rich
star being the mass donor to the accreting WD. The high mass loss
rate demanded by Eq.~(\ref{Eq_mlr}) suggests that the donor should
be a cool AGB carbon star.
For an AGB star the wind velocity is $v_{\rm wind} \sim 10 \kms$,
so the mass loss rate requirement from Eq.~(\ref{Eq_mlr}) would be
$\sim10^{-3}$M$_{\odot}$. This is the order of magnitude of
the highest $\dot M_{\rm wind}$ values determined observationally
for AGB stars which are a few $\times 10^{-4}$M$_{\odot}$
(e.g., Justtanont et al. 2006).
In cool carbon stars, the bulk of the envelope (and wind)
is still hydrogen, but C$_2$ bands are a prominent feature in their
spectra up to $\sim 5000\K$ (e.g., Barnbaum et al. 1996).
If the photospheric temperature of SCP~06F6 was significantly above
that, the presence of Swan bands would imply carbon abundance somewhat
higher than that typical for AGB carbon stars.

As the explosion originates from an accreting WD, the ejection
would likely form a bipolar structure from the very start.
Also, the ejecta mass would be lower compared to the core collapse case.
Both would make it easier for the wind equatorial belt to survive
the SN blast and to feature a C-rich photosphere
at $60 \AU$ and $\sim 100$d after the explosion.

One can conjecture an even more exotic configuration, where the
WD exceeds Chandrasekhar's mass during a common envelope (CE) event
with its AGB companion. The total envelope mass within $60 \AU$ could
then be comparable to the ejecta mass, making it even easier for
the waist/disk structure in the envelope to withstand the SN blast
and form a relatively stable photosphere of $\sim 6000\K$.
A similar explanation was proposed for the prototypical
Ia/IIn SN 2002ic by Livio \& Riess (2003) who argued that it
would favor the double-degenerate SN scenario.
Ofek et al. (2007) suggest a CE connection to deal with the immense
pre-SN mass loss rate of SN 2006gy.
In any case, a binary companion in a natural way provides a bipolar
geometry allowing for simultaneous cool photosphere of the donor wind
and X-rays from the shocked bipolar ejecta
(A different source of X-ray form a SN Ia was studied by Shigeyama et al. 1993).

Normal SNe Ia have the IR lightcurve decline timescales similar to
SCP~06F6 -- in particular they can decline by $\sim 3$~mag in the IR
bands within 100d of the IR maximum, as did SCP~06F6.
The initial brightness rise of the SNe is a few times faster, but
in the scenario considered for the transient, it is the role of the
dense envelope to slow down the initial lightcurve evolution.
The maximum light peak would be smeared out by the interaction
with the envelope, introducing the observed symmetry, the timescale
($\sim 200$d), and the amount of brightness decline.
Despite longer duration of the peak, additional energy deposited in
the opaque envelope by the trapped $\gamma$ photons and by the embedded
collision shocks can supply comparable (or even larger) amplitude
of the optical peak as in a non-interacting case.
The change from the photospheric to nebular phase of a SN
will happen later than normal,
and from that point on the lightcurve would
fall down onto the radioactively powered exponential decline.
The decline rate could resemble that of core collapse SNe
because of the larger than normal for SN Ia total mass in the envelope,
but the levels of the radioactive decay powered luminosity would be
smaller because of the small mass of Ni ejected.

The SN Ia-like scenario does not require a high mass progenitor,
hence it can occur in a faint galaxy with no current star formation.
The initial (main sequence) masses to produce a carbon star on the AGB
lie in the range of ca.~1--$4 M_\odot$
(e.g. Lebzelter \& Hron 2003; Stancliffe et al. 2005),
which translates into main sequence lifetimes of a few
$\times 10^{8}$ to $10^{10}$ yrs.

% ======================================
\subsection{IMBH-WD Tidal Disruptions}
% ======================================

A suggestion that SCP~06F6 might be a star disrupted by a BH was raised by
Gaensicke et al. (2008; also Glennys R. Farrar, private communication 2008).
The observed timescale of the event, the carbon-dominated
spectrum, and the lack of a bright host galaxy lead us to consider
a WD disruption by an IMBH,
as recently modeled numerically by Rosswog et al.\ (2008b).
We bring here a brief description of the tidal destruction of a WD by an IMBH.
Future studies aiming at finding the light curve of such an event
will have to take into account many other processes not considered here
(e.g., Frolov et al. 1994; Wiggins \& Lai 2000; Wilson, J. R. \& Mathews, G. J., 2004;
Brassart \& Luminet 2008; Guillochon et al. 2008).

A crucial parameter in star-BH collisions is
the penetration factor, $\beta = R_{\rm t}/R_{\rm p}$, describing
how close the two
objects get, i.e., how the pericenter of the orbit, $R_{\rm p}$,
compares to the tidal radius at which the tidal acceleration
overcomes stellar (here -- WD) self-gravity,
$ R_{\rm t} = \left({M_{\rm BH}}/{M_{\rm WD}}\right)^{1/3} R_{\rm WD}$.
Using typical CO WD mass of $M_{\rm WD}=0.6 M_\odot$ and
radius of $R_{\rm WD}=9 \times 10^8 \cm$, and an IMBH mass of
$M_{\rm BH}=10^4 M_\odot$,
the tidal radius is:
\begin{equation}
R_{\rm t} = 0.33 \left(\frac{M_{\rm BH}}{10^4 M_\odot}\right)^{1/3}
\left(\frac{M_{WD}}{0.6 M_\odot}\right)^{-1/3}
\left(\frac{R_{WD}}{9\times10^8\cm}\right) R_\odot.
\label{eq:Rp2}
\end{equation}
Tidal disruption occurs for $\beta \ga 1$.
But for a fixed $\beta$, a WD will only be tidally
disrupted by a BH within a certain mass range (Luminet \& Pichon
1989a,b; Rosswog et al. 2008b).
If the BH is too massive, the WD is swallowed
before being tidally disrupted, with no significant emission.
If the BH is too low-mass, it enters
the WD instead of disrupting it.
For too high values of the penetration factor there is no
chance for a tidal disruption event at all, because the
strongly relativistic regime (the star enters the BH/the BH enters the star)
would come into play at any mass.
For intermediate masses and moderate values of $\beta$,
the interaction with the BH compresses the WD to a pancake-like shape
and tears it apart.
Part of the material will be gravitationally bound to the IMBH
and will form an accretion disk
which can supply
accretion luminosity close to the Eddington luminosity,
mainly in soft X-rays,
for about a year following the disruption (Rosswog et al. 2008b),
similar to the $\sim 200$~days duration of the SCP~06F6 event.
Explosive thermonuclear ignition of the compressed WD material is also possible
and Rosswog et al. (2008a) proposed this as a mechanism for
a new type of underluminous SNe, later suggesting that SCP~06F6 can be the first
discovered object of this kind (Rosswog et al. 2008b,c).
But since SCP~06F6, if at $z=0.143$, was clearly not underluminous compared
to SNe, and accretion alone can supply enough luminosity, we do not
consider a thermonuclear explosion as a necessary ingredient in this case.

The optical observations do, however, require another element: enough matter
around the object to form a relatively stable photosphere of $\sim 6000\K$
at $60\AU$. We postulate that this can be due to a strong wind from the
accretion disk, colliding with the matter ejected at earlier phases of the
disruption.
Note that in this case the X-ray emission can come either from a shocked
fast outflow material or from the inner parts of the accretion disk,
in fact similar to what would occur in the WD-asteroid model.

According to Luminet \& Pichon (1989b),
the maximum BH mass still allowing tidal disruption of a $0.6 M_\odot$ WD is
$M_{\rm BH,lim}=3\times10^4 M_\odot$.
However, such a massive BH could only exert weak tidal effects without
swallowing the WD.
The optimal BH mass, allowing the strongest tidal effects, is
$M_{\rm BH,opt}=1.53\times10^3 M_\odot$.
For SCP~06F6, the minimum required BH mass
can be estimated by comparing the BH Eddington luminosity
to the SCP~06F6 luminosity.
The limit turns out to be
$M_{\rm BH} \ga 2 \times 10^3 M_\odot$
when considering only the optical peak of
$L_{\rm p} \simeq 1.3 \times 10^{42} \erg \s^{-1}$
with $\kappa \simeq 0.1 \cm^2 \g^{-1}$ as before,
or $M_{\rm BH} \ga 2 \times 10^4 M_\odot (\kappa/0.2 \cm^2 \g^{-1})$
when including the X-ray emission (with $\kappa$ scaling appropriate for
high temperatures).
Exclusion of X-rays is justified if they originate in the post-shock gas of
a fast outflow.
With all these considerations, the BH mass range relevant
for the IMBH-WD scenario for SCP~06F6 is a few $\times 10^3$ to
a few $\times 10^4 M_\odot$.

Accounting for the Swan spectrum requires a considerable amount of carbon,
either coming from the initial WD composition
or produced during the interaction with the IMBH.
The former is the case for the CO WD with $\beta=1.5$ from Rosswog et al.:
after the disruption the material retains its initial composition
(assumed in their simulation to be 50\% oxygen and 50\% carbon).
Even igniting CO WDs keep carbon-rich external layers.
The latter could occur for a helium WD, when the tidal compression
triggers a thermonuclear explosion.
While for a strongly interacting He WD the final nucleosynthesis products
are dominated by silicon and iron (Rosswog 2008b),
carbon can be the dominant end-product in a weaker interaction,
when the compression phase is
too short to fuse heavier elements (Luminet \& Pichon 1989b).
Note that such interrupted burning
would produce an even more underluminous SN.

In the standard disruption models,
the mass accretion rate (and hence the accretion luminosity)
falls off at late times as $\sim t^{-5/3}$ or slower,
where $t=0$ is set at the maximum of the interaction strength,
much before the
lightcurve reaches its maximum (Lodato et al. 2008; Ramirez-Ruiz \& Rosswog 2008).
For SCP~06F6, there are very few measurements at late times, and their errorbars are quite large.
However, it is clear that with $t=0$ at the beginning of the SCP~06F6 event,
the late decline rate is faster than $t^{-5/3}$ both in the optical and
in the X-rays. The X-ray signal dropped over two orders of magnitude within
$\sim 100$d.
This again can be reconciled if a strong wind from the accretion disk
depletes the disk in a few months and terminates the backflow onto the BH.
Another possibility is that at late times a large fraction of the accreted
energy goes into launching jets.

The outer radius of the disk is $R_{\rm disk} \simeq \textrm{few} \times R_{\rm t}$
(Rosswog et al. 2008b).
This allows a simple estimate of
the outflow terminal speed, influenced by the interaction
with material residing around and still falling in:
\begin{equation}
v_{\rm wind} =\xi v_{\rm esc}(R_{\rm disk})
\simeq 6\times10^3 \left(\frac{\xi}{0.1}\right)
\left(\frac{M_{\rm BH}}{10^4 M_\odot}\right)^{1/3}\kms
\label{eq:vwind}
\end{equation}
where the scaling factor $\xi$ encompasses our uncertainity of the disk size
and of the wind slow-down. From Eq.~(\ref{Eq_mlr}) with  $v_{\rm wind} =
6000 \kms$, the mass loss rate required for a photosphere at
$R_{\rm ph-eg}= 60\AU$ is $\dot M_{\rm wind}
\simeq 0.7 \msyr$.
Over the 200 days eruption, a total mass of a few $0.1 M_\odot$ would
be lost from the accretion disk and its vicinity. The outflow velocity and
mass would resemble those of SN ejecta.

% ======================================
\section{SUMMARY}
% ======================================

We compared four
scenarios for the SCP~06F6 transient event
(Barbary et al. 2008),
one galactic and three extragalactic (with a redshift of $z=0.143$ as
suggested by Gaensicke et al. 2008a):

1. \emph{A Galactic event of an asteroid colliding with a WD},

2. \emph{A new class of core collapse SN}, as suggested by Gaensicke et al. (2008a),

3. \emph{Type Ia SN explosion},
resembling the hybrid Ia/IIn type,

4. \emph{Tidal disruption of a CO WD by an intermediate mass black hole (IMBH)},
as recently studied in detail by
Rosswog et al. (2008b,c) and Lodato et al. (2008).

In principle, all four scenarios can lead to a transient event, and we
expect that such events will eventually be discovered.
However, as we presently attempt to explain SCP~06F6,
we compare the models to the observations of this transient in Table 1
%============================== TABLE ===============================
%\begin{center}
\begin{table}

Table 1: Comparison of the models

\footnotesize
\bigskip
\begin{tabular}{|l||l|l|l|l|}
\hline
Model      & Extragalactic    & Extragalactic   & Extragalactic       & Galactic WD-  \\
           & core collapse SN & SN Ia-like      & IMBH-WD collision   & asteroid merger \\
\hline
Reference  &  Gaensicke       & This paper      & Rosswog             & This paper\\
           & et al. (2008a)   &                 & et al. (2008c)      & \\
\hline
\hline
(1) Spectrum  & \multicolumn{3}{|l|}{Like a redshifted carbon star with absorption Swan bands} & Absorption \\
              & \multicolumn{3}{|l|}{}                               & Swan bands \\
              & \multicolumn{3}{|l|}{}                               & distorted by  \\
              & \multicolumn{3}{|l|}{}                               & magnetic fields \\
\hline
(2) Optical & \multicolumn{2}{|l|}{Within the range of SNe}  & Fits the Eddington      & Determines       \\
Luminosity  & \multicolumn{2}{|l|}{}                         & luminosity of an IMBH   & the colliding    \\
            & \multicolumn{2}{|l|}{}                         & tidally disrupting a WD  & asteroid mass    \\
\hline
(3) Timescale & Typical of SNe,  & Typical of SNe, & Typical accretion        & KH timescale   \\
              & but the event    & some SNe Ia     & time, but the decline    & of merged      \\
              & declines faster  & have similar    & is faster than predicted & product        \\
              & than core        & decline rates   &                          &                \\
              & collapse SNe     &                 &                          &                \\
\hline
(4) Very bright&\multicolumn{2}{|l|}{$\times 100$ more than in typical SNe.} &\multicolumn{2}{|l|}{X-rays from accretion disk} \\
in X-rays      &\multicolumn{2}{|l|}{Requires X-rays to come from}           &\multicolumn{2}{|l|}{and shocked expelled material} \\
               &\multicolumn{2}{|l|}{shocked polar ejecta}                   &\multicolumn{2}{|l|}{} \\
\hline
(5) Undetected &\multicolumn{2}{|l|}{Very faint host galaxy}  & Allowed BH mass     & Very cool WD \\
progenitor     &\multicolumn{2}{|l|}{}                        & range accounts for  & progenitor \\
               &\multicolumn{2}{|l|}{}                        & a very faint galaxy &     \\
\hline
Strong points & a) Peak        & a) Spectrum:    & a) Spectrum: origin    & a) Spectrum:      \\
in explaining & optical        & carbon from     & of carbon is explained & origin of carbon  \\
SCP~06F6      & luminosity     & companion's     & b) Luminosity          & is explained      \\
              & b) Timescale   & dense wind      & c) Time scale          & b) Luminosity     \\
              &                & b) Peak         & d) Strong X-rays       & fits an asteorid  \\
              &                & optical         & e) If BH is at         & c) Timescale      \\
              &                & luminosity      & the nucleus the model  & d) Strong X-rays  \\
              &                & c) Timescale    & accounts for a very    &                   \\
              &                & d) Faint galaxy & faint galaxy           &                   \\
              &                & is possible     & & \\
\hline
Weak  points  & a) Spectrum:        & a) Mass loss    & a) Observed decline is                 & a) Spectrum:    \\
in explaining & C-rich envelope     & rate extreme    & steeper than predicted                 & very strong     \\
SCP~06F6      & hard to explain     & for an AGB      & b) Difficult to account                & magnetic fields \\
              & b) Strong X-rays    & star            & for a $60 \AU$ ($\sim 10^5 R_{\rm g}$) & are required    \\
              & c) No star-         & b) Special      & photosphere                            & to explain      \\
              & forming galaxy      & geometry        & {\it Note: An addition of}             & the location    \\
              & is observed         & is required     & {\it a disk wind solves}               & of Swan bands   \\
              & d) Special          &                 & {\it these problems}                   &       \\
              & geometry            &                 &                                        &       \\
              & is required         &                 &                                        &       \\
\hline
\end{tabular}

\footnotesize
\bigskip

\normalsize
\end{table}
%\end{center}
% ======================== END TABLE ===============================
with the following comments (numbers as in the first column of the table):
\newline
(1) Redshifted carbon star-spectrum does better than what we can find
for a Galactic source. A Galactic source requires absorption Swan bands
distorted by strong magnetic fields, possibly with addition of some emission Swan
bands,
which also puts its distance to be $\sim 1.5 \kpc$.
Despite its complexity, we cannot rule out such a Galactic source.
The IMBH-WD disruption model with no nuclear burning, i.e., a CO WD seems to
do better in accounting for the spectrum.
A SN embedded in C-rich matter is also fine, but the origin of carbon
needs explanation. The Ia-like SN scenario (SN Ia/IIn) seems more natural
than core collapse,
which would require a very unusual binary interaction.
\newline
(2) For the Galactic source, total emitted energy is used to constrain the
mass of the asteroid.
For the extragalactic sources, the luminosity is within the range
of SNe.
For the IMBH-WD disruption model there is as yet no constraint from theory or
observation on the luminosity.
\newline
(3) As far as timescales, the SN models have the closest match,
although the observed decline is faster than those of most core collapse SNe.
Some SNe type IIn have equally fast decline in the IR bands
and the SN Ia-like model (Ia/IIn) does good as well.
The WD-asteroid model (where the event timescale is the Kelvin-Helmholtz
timescale),
were not observed in the past, but from theoretical point of view can
explain the timescale.
As for the IMBH-WD tidal disruption model, we must incorporate a new ingredient to existing
calculations of the disruption process, which we suggest is a strong disk wind.
\newline
(4) In the Galactic source the X-rays originate from close to the WD, as in
dwarf novae. This can be the X-ray source at late times, when the space above and below the disk
has been cleared from dense gas. Another possibility is the ejection of jets (or a collimated fast wind),
which when collide with the circum-system material become the X-ray source.
The same two mechanisms can operate in the IMBH-WD tidal interaction model.
It seems that the SNe models should also employ a non-spherical geometry that
allows fast polar ejecta to be shocked and emit X-rays.
This further disfavors a core collapse SN based on a single star model; we can
safely conclude that if SCP~06F6 is a core collapse SN, its peculiarity
comes from a strong binary interaction.
We recall that the peculiarity in the SN Ia-like model comes from a carbon star
with a high mass loss rate, that also transfers mass to the pre-exploding WD.
Namely, in the SNe models of SCP~06F6 we attribute the peculiarity to a rare binary companion.
\newline
(5) The non detection of a galaxy at the location of the event is more of a
problem for core collapse SN models because the galaxy is fainter than known star-forming
galaxies.
Less of a problem to SN Ia-like, and even less of a problem for the IMBH-WD
disruption model, as a galaxy with no recent star formation can do with these two models.
Moreover, if we assume that the IMBH is the BH at the nucleus of the
galaxy, then a very faint dwarf galaxy is indeed expected.
The non detection is not a problem for the WD-asteroid merger model, as it
is used to constrain the progenitor distance to $D >1.2  \kpc$ (Barbary et al. 2008).

We summarize the strong and weak points of the four models in the last two rows of Table 1.
The most important thing to note is that
IMBH-WD tidal disruption model has weak points if we use the existing numerical simulations
for this process as they are.
However, we can solve these problems if we add a very strong disk outflow at velocities of
${\rm several} \times 10^3 \km \s^{-1}$.
 (i) \emph{Photosphere}: With a mass of $0.1-0.3 M_\odot$ being lost at $3000-10^4 \kms$,
we can account for the photosphere (defined at $\tau=2/3$) at tens of AU, and in particular
of $60 \AU$ at maximum light.
(ii) \emph{Lightcurve}: With the rapid removal of mass from the disk, and the interaction
of this wind with inflowing gas, we can account for
the lightcurve declining rate that is much more rapid than $t^{-5/3}$.
This outflow is powered by accretion close to the Eddington limit, and
possibly by extreme amplification of magnetic fields in the accretion disk.

For that, our favorite model is the IMBH-WD model.
Second we rank the type Ia-like SN model, with a carbon star mass donor
companion. Third we rank the WD-asteroid Galactic model.
The
core collapse model comes last, as it seems it must be very
peculiar, and involve some kind of strong binary interaction.

The predictions of the different models are straightforward,
allowing for an observational test.
The Galactic scenario implies that a faint WD, with a possible IR excess,
resides at the event location.
The extra-galactic scenarios predict
a very faint galaxy in that direction.
The type II SN model implies the presence of a recently active star forming region.

\acknowledgments
We thank Glennys R. Farrar, Stephan Rosswog, Enrico Ramirez-Ruiz and Giuseppe Lodato
for helpful discussions.
This research was supported in part by a grant from the Israel Science
Foundation, and from the the Asher Space Research Institute in the Technion.


\begin{references}

\reference{} Adelman-McCarthy, J., et al. 2008, ApJS, 175, 297

\reference{} Aldering, G., et al. 2006, ApJ 650, 510  %more than 10 authors

\reference{} Alexander, D. R., \& Ferguson, J. W. 1994, ApJ, 437, 879

\reference{} Angel, J. R. 1977, ApJ, 216, 1

\reference{} Barbary, K., et al. 2008, accepted to ApJ (arXiv:0809.1648)

\reference{} Barnbaum, C., Stone, R. P. S., Keenan, P. C. 1996, ApJS, 105,
     419

\reference{} Becklin, E. E., Farihi, J., Jura, M., Song, I., Weinberger,
     A. J., \& Zuckerman, B. 2005, ApJ 632, L119

\reference{} Brassart, M. \& Luminet, J.P. 2008, A\&A, 481, 259

\reference{} Bues, I. 1999, in ASP Conf. Ser. 169, 11th European Workshop
   on White Dwarfs, ed. J. E. Solheim \& E. G. Meistas (San Francisco: ASP),
   240

\reference{} Chugai, N. N., Blinnikov, S. I., Cumming, R. J.,
     Lundqvist, P., Bragaglia, A., Filippenko, A. V., Leonard, D. C.,
     Matheson, T., \& Sollerman, J. 2004, MNRAS 352, 1213

\reference{} Chugai, N. N., Danziger, I. J., 2003, AstL, 29, 649

\reference{} Contardo, G., Leibundgut, B., \& Vacca W. D., 2000, A\&A, 359,
   876

\reference{} Corradi, R. L. M., \& Munari, U. eds., 2007, `The Nature of V838 Monocerotis
     and its Light Echo' (San Francisco, CA: ASP)

\reference{} Crowther, P. A. 2007, ARA\&A, 45, 177

\reference{} Debes, J. H., \& Sigurdsson, S. 2002, ApJ, 572, 556

\reference{} Deng, J., Kawabata, K. S., Ohyama, Y., Nomoto, K., Mazzali, P. A., Wang, L.,
    Jeffery, D. J., Iye, M., Tomita, H., \& Yoshii, Y. 2004, ApJ, 605, L37

\reference{} de Freitas Pacheco, J. A., Singh, P. D., \& Landaberry, S. J. C.
     1988, MNRAS, 235, 457

\reference{} Dufour, P., Bergeron, P., \& Fontaine, G. 2005, ApJ, 627, 404

\reference{} Dufour, P., Fontaine, G., Liebert, J., Schmidt, G. D., \&
     Behara, N. 2008, ApJ, 683, 978

\reference{} Ferguson, J. W., Alexander, D. R., Allard, F., Barman, T.,
     Bodnarik, J. G.,  Hauschildt, P. H., Heffner-Wong, A., \& Tamanai, A. 2005, ApJ, 623, 585

\reference{} Filippenko, A. V., 1997, ARA\&A, 35, 309

\reference{} Fink, U., \& Hicks, M. D. 1996, ApJ 459, 729

\reference{} Frolov, V. P., Khokhlov, A. M., Novikov, I. D., \& Pethick, C. J. 1994, ApJ, 432, 680

\reference{} Gaensicke, B. T., Koester, D., Marsh, T. R., Rebassa-Mansergas, A.,
     \& Southworth, J. 2008b (arXiv:0809.2600)

\reference{} Gaensicke, B. T., Levan, A. J., Marsh, T. R., \& Wheatley, P.J.
     2008a (arXiv:0809.2562)

\reference{} Gaensicke, B. T., Marsh, T. R., Southworth, J., \&
     Rebassa-Mansergas, A., 2006, Sci, 314, 1908

\reference{} Gaensicke, B. T., Marsh, T. R., \& Southworth, J. 2007, MNRAS,
     380, L35

\reference{} Gal-Yam, A., Leonard, D. C., Fox, D. B., Cenko, S. B., Soderberg, A. M., Moon, D.-S.,
    Sand, D. J., Li, W., Filippenko, A. V., Aldering, G., \& Copin, Y. 2007, ApJ, 656, 372

\reference{} Garc\'{i}a-Segura, G.; Mac Low, M.-M.; Langer, N. 1996, A\&A, 305,
     229

\reference{} Graham, J. R., Matthews, K., Neugebauer, G., \& Soifer, B. T.
     1990, ApJ, 357, 216

\reference{} Guillochon, J., Ramirez-Ruiz, E. \&  Rosswog, S. 2008 (arXiv:0811.1370)

\reference{} Hall, P. B., \& Maxwell, A. J. 2008, ApJ, 678, 1292

\reference{} Hamuy, M. et al. 2003, Nature, 424, 651  %more than 10 authors

\reference{} Ho, W. C. G., Van Dyk, S. D., Peng, Ch. Y., Filippenko, A. V.,
     Leonard, D. C., Matheson, T., Treffers, R. R., Richmond, M. W. 2001,
     PASP, 113, 1349

\reference{} Hoffman, J. L. 2007, in: Supernova 1987A: 20 Years After:
     Supernovae and Gamma-Ray Bursters, eds. S. Immler, K. W. Weiler,
     \& R. McCray, (New York: AIP), AIP Conf. Proc. 937, 365

\reference{} Iben, I. Jr., Tutukov, A, V., \& Yungelson, L. R. 1996, ApJ, 456, 750

\reference{} Jura, M. 2003, ApJ 584, L91

\reference{} Jura, M., Farihi, J., Zuckerman, B., \& Becklin, E. E. 2007,
     AJ, 133, 1927

\reference{} Justtanont, K., Olofsson, G., Dijkstra, C., \& Meyer, A. W.
    2006, A\&A, 450, 1051

\reference{} Kami\'{n}ski, T. 2008, A\&A, 482, 803

\reference{} Kawka A., Vennes S., Schmidt G. D., Wickramasinghe D. T., \&
     Koch R. 2007, ApJ, 654, 499

\reference{} Kilic, M., von Hippel, T., Leggett, S. K., \& Winget, D. E.
     2006, ApJ, 646, 474

\reference{} Koester, D., Provencal, J., \& Shipman, H. L. 1997, A\&A, 230, L57

\reference{} Kotak, R., Meikle, W. P. S., Adamson, A., Leggett, S. K. 2004,
     MNRAS, 354, L13

\reference{} Kotak, R., Vink, J. S. 2006, A\&A, 460, L5

\reference{} Krishna Swamy, K. S. 1997, Physics of Comets, Singapore: World
   Scientific, 2nd ed.

\reference{} Leahy, D., \& Ouyed, R. 2008, MNRAS, 387, 1193

\reference{} Lebzelter, T., \& Hron, J. 2003, A\&A, 411, 533

\reference{} Liebert, J., Angel, J. R., Stockman, H. S., \& Beaver, E. A.
     1978, ApJ, 225, 181

\reference{} Livio, M., Riess, A. G. 2003, ApJ, 594, 93

\reference{} Lodato, G., King, A. R., \& Pringle, J. E. 2008 (arXiv:0810.1288)

\reference{} Luminet, J. P., \& Pichon, B. 1989a, A\&A, 209, 85

\reference{} Luminet, J. P., \& Pichon, B. 1989b, A\&A, 209, 103

\reference{} Marigo, P. 2002, A\&A, 387, 507

\reference{} Nomoto, K., Tominaga, N., Umeda, H., Kobayashi, Ch., \&
     Maeda, K. 2006, Nuc. Phys. A, 777, 424

\reference{} O'Dell, C. R., Robinson, R. R., Krishna Swamy, K. S.,
     McCarthy, P. J., Spinrad, H. 1988, ApJ, 334, 476

\reference{} Ofek, E. O., et al. 2007, ApJ, 659, L13

\reference{} Paczy\'{n}ski, B., 1998, in: Gamma-Ray Bursts: 4th Huntsville
    Symposium, eds. Ch. A. Meegan, R. D. Preece, \& T. M. Koshut,
    (Woodbury: AIP), AIP Conf. Proc., 428, 783

\reference{} Pastorello, A., Ramina, M., Zampieri, L., Navasardyan, H.,
     Salvo, M., \& Fiaschi, M. 2005, in: Cosmic Explosions, On the 10th
     Anniversary of SN1993J. Proceedings of IAU Colloquium 192, eds. J.M.
     Marcaide \& K. W. Weiler, (Berlin: Springer), Springer Proceedings
     in Physics, 99, 195

\reference{} Ramirez-Ruiz, E. \& Rosswog, S. 2008 (arXiv:0808.3847)

\reference{} Rao N. K., \& Lambert, D. L. 1993, AJ, 105, 1915

\reference{} Richardson, D., Branch, D., Casebeer, D., Millard, J.,
     Thomas, R. C., Baron, E. 2002, AJ, 123, 745

\reference{} Rosswog, S., Ramirez-Ruiz, E. \& Hix, R. 2008a, ApJ, 679, 1385

\reference{} Rosswog, S., Ramirez-Ruiz, E. \& Hix, R. 2008b (arXiv:0808.2143)

\reference{} Rosswog, S., Ramirez-Ruiz, E., \& Hix, W. R. 2008c (arXiv:0811.2129)

\reference{} Schlegel E. M., 1990, MNRAS, 244, 269

\reference{} Schmidt, G. D., Bergeron, P., Fegley, B. Jr., 1995, ApJ, 443, 274

\reference{} Schmidt, G. D., Liebert, J., Harris, H. C., Dahn, C. C.,
Leggett, S. K., 1999, ApJ, 512, 916

\reference{} Shara, M. M. 2002, ASPC, 263, 1S

\reference{} Shigeyama, T., Kumagai, S., Yamaoka, H., Nomoto, K.,
     \& Thielemann, F.-K. 1993, A\&AS, 97, 223

\reference{} Silber, A., Vrtilek, S. D.,l \& Raymond, J. C. 1994, ApJ, 425, 829

\reference{} Smith, N. 2008, in IAU Symp. 250, Massive Stars as Cosmic
     Engines, eds. F. Bresolin, P. A. Crowther, \& J. Puls,
     (Cambridge: Cambridge Univ. Press), 193

\reference{} Smith, N., et al. 2007, ApJ, 666, 1116  %more than 10 authors

\reference{} Smith, N., Chornock, R., Li, W., Ganeshalingam, M.,
     Silverman, J. M., Foley, R. J., Filippenko, A. V., Barth, A. J. 2008,
     ApJ, 686, 467

\reference{} Smith, N., McCray, R. 2007, ApJ, 671, L17

\reference{} Smith, N., Owocki, S. P. 2006, ApJ, 645, L45

\reference{} Soker, N., \& Tylenda, R. 2003, ApJ, 582, L105

\reference{} Soker, N., \& Tylenda, R. 2006, MNRAS, 373, 733

\reference{} Sparks, W. B., et al. 2008, AJ, 135, 605 % more than 10 authors

\reference{} Stancliffe, R. J., Izzard, R. G., \& Tout, Ch. A., 2005, MNRAS,
   356, L1

\reference{} Stawikowski, A., \& Swings, P. 1960, AnAp, 23, 585

\reference{} Trundle, C., Kotak, R., Vink, J. S., \& Meikle, W. P. S. 2008,
     A\&A, 483, L47
\reference{} Tylenda, R. 2005, A\&A, 436, 1009

\reference{} Tylenda, R. \& Soker, N., 2006, A\&A, 451, 223

\reference{} van Dyk, S. D., Weiler, K. W., Sramek, R. A., Schlegel, E. M.,
     Filippenko, A. V., Panagia, N., Leibundgut, B. 1996, AJ, 111, 1271

\reference{} Vanlandingham, K. M., et al. 2005, AJ, 130, 734 % more than 10 authors

\reference{} van Teeseling, A., Beuermann, K., Verbunt, F. 1996, A\&A, 315, 467

\reference{} von Hippel, T., Kuchner, M. J., Kilic, M., Mullally, F., Reach, W. T.
    2007, ApJ 662, 544

\reference{} Wells, L. A., et al. 1994, AJ, 108, 2233 % more than 10 authors

\reference{} Wheeler, J. C., Harkness, R. P., Khokholv, A. M., \& Hoeflich,
    P. 1995, PhRep 256, 211

\reference{} Wickramasinghe, D. T., \& Ferrario, L. 2000, PASP, 112, 873

\reference{} Wiggins \& Lai 2000, ApJ, 532, 530

\reference{} Wilson, J. R. \& Mathews, G. J., 2004, ApJ, 610, 368W

\reference{} Wisniewski, J. P., Clampin, M., Bjorkman, K. S., Barry, R. K.
    2008, ApJ, 683, L171

\reference{} York, D. G., et al. 2000, AJ, 120, 1579

\reference{} Zuckerman, B., \& Becklin, E. E. 1987, Nature 330, 138

\reference{} Zuckerman, B., Koester, D., Melis, C., Hansen, B., \& Jura, M.
    2007, ApJ

\reference{} Zuckerman, B., \& Reid, I. N. 1998, ApJ, 505, L143

\end{references}
\end{document}